\documentclass{article}
\usepackage{dcase2020,amsmath,graphicx,url,times,booktabs, tabularx}
\usepackage{amssymb, bm}
\usepackage{calrsfs}
\usepackage{color}
\usepackage[utf8]{inputenc}
\usepackage{mathrsfs, mathtools, bbm}
\usepackage[final]{microtype}
\usepackage{booktabs}
\usepackage{array}
\usepackage{enumitem}
\usepackage{soul}
\usepackage{cite}
\usepackage{siunitx}
\graphicspath{{./figs/}}

\DeclareMathAlphabet{\pazocal}{OMS}{zplm}{m}{n}
\newcolumntype{P}[1]{>{\centering\arraybackslash}p{#1}}

\title{Event-Independent Network for Polyphonic \\Sound Event Localization and Detection}

\name{Yin Cao,$^{1}$ Turab Iqbal,$^{1}$ Qiuqiang Kong,$^{2}$ Yue Zhong,$^{1}$ Wenwu Wang,$^{1}$ Mark D. Plumbley$^{1}$}
\address{$^{1}$Centre for Vision, Speech and Signal Processing (CVSSP), University of Surrey, UK \\
\{yin.cao, t.iqbal, y.zhong, w.wang, m.plumbley\}@surrey.ac.uk \\
$^{2}$ByteDance Shanghai, China\\ kongqiuqiang@bytedance.com\\}

\begin{document}

\ninept
\maketitle

\begin{sloppy}

\begin{abstract}
Polyphonic sound event localization and detection is not only detecting what sound events are happening but localizing corresponding sound sources. This series of tasks was first introduced in DCASE 2019 Task 3. In 2020, the sound event localization and detection task introduces additional challenges in moving sound sources and overlapping-event cases, which include two events of the same type with two different direction-of-arrival (DoA) angles. In this paper, a novel event-independent network for polyphonic sound event localization and detection is proposed. Unlike the two-stage method we proposed in DCASE 2019 Task 3, this new network is fully end-to-end. Inputs to the network are first-order Ambisonics (FOA) time-domain signals, which are then fed into a 1-D convolutional layer to extract acoustic features. The network is then split into two parallel branches. The first branch is for sound event detection (SED), and the second branch is for DoA estimation. There are three types of predictions from the network, SED predictions, DoA predictions, and event activity detection (EAD) predictions that are used to combine the SED and DoA features for on-set and off-set estimation. All of these predictions have the format of two tracks indicating that there are at most two overlapping events. Within each track, there could be at most one event happening. This architecture introduces a problem of track permutation. To address this problem, a frame-level permutation invariant training method is used. Experimental results show that the proposed method can detect polyphonic sound events and their corresponding DoAs. Its performance on the Task 3 dataset is greatly increased as compared with that of the baseline method.
\end{abstract}


\section{Introduction}
\label{sec:intro}
Sound event localization and detection (SELD) has become a more and more popular research topic since DCASE 2019. It detects types of sound events and localizes corresponding sound sources. This year, DCASE 2020 Task 3 \cite{politis2020dataset, Adavanne2018_JSTSP, Mesaros_2019_WASPAA, DCASE2020Task3} introduces additional challenges in moving sources and polyphonic cases that include the same class of event but with different direction-of-arrival (DoAs).

For DCASE 2019 Task 3, we introduced a two-stage method for polyphonic SELD \cite{cao2019polyphonic}. Although we obtained a good ranking, the method was not designed as an actual polyphonic localization method for the reason that it lacks the ability to detect sound events of the same type but with different DoAs. Besides, it is not an elegant end-to-end system, which needs to be trained for two steps.

In this paper, we propose a redesigned event-independent end-to-end system for polyphonic SELD. It is designed for overlapping-event cases, including the presence of the same type of event with different DoAs. It is also convenient to expand the system to the case of more than two overlapping events. The source code is released on GitHub\footnote{\url{https://github.com/yinkalario/EIN-SELD}}. Our contributions are three-fold. 1) The proposed system predicts overlapping events using track-wise outputs, that is, it predicts event and corresponding DoA for each track. Within each track there may only be maximally one event and corresponding DoA existing. 2) Frame-level permutation invariant training is adopted to solve the track permutation problem. 3) An event activity detection (EAD) prediction is added to combine sound event detection (SED) and DoA estimation feature embeddings to predict the on-set and off-set times more accurately.

\begin{figure}[ht]
  \centering
  \scalebox{.9}{\centerline{\includegraphics[width=\columnwidth]{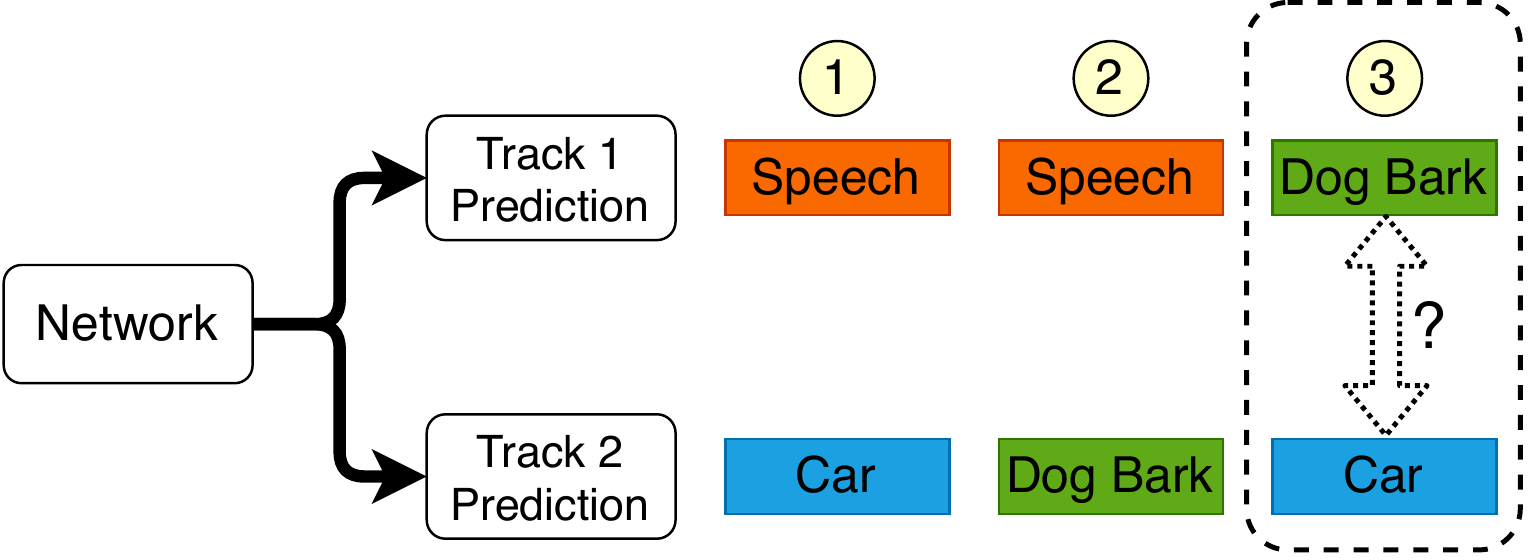}}}
  \caption{Illustration of the track permutation problem. Numbers mean different group of labels.}
  \label{fig:track_permutation}
\end{figure}

We adopt the track-wise prediction to tackle the overlapping-event scenarios. It is needed to pre-determine the number of tracks according to the maximum number of overlapping events. These tracks are event-independent, which means the prediction on each track can be of any type of event. They can even be the same type of event which indicates that two same-type events with different DoAs are predicted. It is also reasonable to assume these tracks are event-independent. Consider a polyphonic prediction case illustrated in Fig. \ref{fig:track_permutation}. The network has one prediction for each track, within which there could only be maximally one event and corresponding DoA. There are three groups of labels which are all potentially two-event overlapping cases. It is assumed that, for the first group, the ``speech" label and the ``car" label are tied to track 1 and 2. For the second group, it is reasonable to still assign the ``speech" label to track 1, and the new ``dog bark" label to track 2. However, for the third group of labels, it is hard to decide to which track the ``dog bark" or the ``car" label should be assigned. In other words, track permutation problems emerge if track-wise predictions are used.

To address the track permutation problem, frame-level permutation invariant training (denoted as tPIT) is used. The frame-level permutation invariant training was first proposed for speaker-independent source separation \cite{yu2017permutation, kolbaek2017multitalker}. For our SELD problem, the tPIT is implemented by examining all possible track permutations in each frame during training, and then the lowest frame-level loss is selected among these track permutations for the backward propagation to train the model. In this way, the optimal local assignment of track-event pairs can be reached, thus leading to the excellent SED and DoA prediction performance frame-wise.

In order to estimate the frame-level information more accurately, the features from both SED and DoA branches are combined to predict event activities. The aim of this EAD is to constrain the detection of the existence of events (or DoAs) not only from both SED and DoA features. That means SED and DoA predictions have a mutual dependence rather than a one-way dependence. With the proposed system, experimental results show that the performance is greatly increased compared with that of the baseline system.

The rest of the paper is arranged as follows. In Section \ref{sec:method}, the proposed learning method is described in detail, including features, network architecture and permutation invariant training. Experimental results and discussions are shown in Section \ref{sec:experiments}. Finally, conclusions are summarized in Section \ref{sec:conclusion}.

\section{Related works}
\label{sec:related_works}

\subsection{Sound Event Localization and Deteciton}

Sound event localization and detection is a novel topic that has wide applications \cite{virtanen2018computational}. In DCASE 2019 Task 3 \cite{Adavanne2019_DCASE, Adavanne2018_JSTSP}, the TAU Spatial Sound Events 2019 dataset was released \cite{Mesaros2016_MDPI}. There were several innovative papers based on that dataset. Mazzon et al. proposed a spatial-augmentation method which rotates and reflects sources so that unseen DoA data and corresponding labels are produced \cite{Mazzon2019}. Grondin et al. used a CRNN on pairs of microphones to localize and detect sound events \cite{Grondin2019}. Sundar et al. proposed an encoding scheme to represent spatial coordinates of multiple sources \cite{sundar2020raw}. Although the two-stage method we proposed \cite{cao2019polyphonic} last year achieved the second-best performance in DCASE 2019 Task 3, it is not an elegant end-to-end system and is not designed for overlapping events of the same type with different DoAs. Then, an idea of similar track-wise prediction was proposed by Nguyen et al. \cite{nguyen2020sequence}. However, their system did not show a reasonable bond between SED and DoA predictions, they assume the track prediction with the highest probability of SED corresponds to the same highest probability track of DoA. In this paper, it will be shown in the following part that the system proposed is a more complete system in this sense.

\subsection{Permutation Invariant Training}

Permutation invariant training (PIT) was first proposed to tackle the problem of speaker-independent multi-talker speech separation \cite{yu2017permutation}, commonly known as the cocktail-party problem. PIT combines the label assignment and minimization together, and can be implemented inside the network structure. It first assigns the best predict-target pairs according to which way the total loss is the smallest, and then minimizes the loss given the assignment. PIT was then extended to frame-level and utterance-level PIT \cite{kolbaek2017multitalker, wang2018multi, xu2018single, fan2018utterance}, which were utilized by a range of speaker-independent speech separation researchers \cite{liu2019divide, luo2018speaker, liu2020deep, ephrat2018looking, chen2017deep}. In our proposed method, a track-wise output format is used. Tracks are event-independent, that is, tracks can predict any type of event. This generates a problem of track permutation. It will be shown that by adopting a similar idea to the frame-level PIT, the track permutation problem can be excellently solved.

\section{The method}
\label{sec:method}

The proposed method is described in this section. Features used are logmel and intensity vector. They are calculated inside the network using a 1-D convolutional layer. The network architecture and permutation invariant training will be introduced in detail.

\subsection{Features}
\label{ssec:features}
In this paper, a logmel spectrogram feature is used for SED, while an intensity vector from fisrt-order ambisonics (FOA) in logmel space is used for DoA estimation. These features are directly calculated using a 1-D convolutional layer instead of being pre-calculated off-line.

FOA, which is also known as B-format, includes four channels of signals, \(\mathrm{w}, \mathrm{x}, \mathrm{y}\) and \(\mathrm{z}\). These four channel signals indicate omni-directional, $x$-directional, $y$-directional and $z$-directional components, respectively. The instantaneous sound intensity vector can be expressed as $\boldsymbol{I}=p \mathbf{v}$, where $p$ is the sound pressure and can be obtained with $\mathrm{w}$, $\mathbf{v}=\left(\mathbf{v}_{x}, \mathbf{v}_{y}, \mathbf{v}_{z}\right)^{\mathrm{T}}$ is the particle velocity vector and can be estimated using $\mathrm{x}, \mathrm{y}$ and $\mathrm{z}$. An intensity vector carries the information of the acoustical energy direction of a sound wave, its inverse direction can be interpreted as the DoA, hence the FOA-based intensity vector can be directly utilized for DoA estimation \cite{kotus2014detection, ahonen2007teleconference, Cao2019, perotin2019crnn}.

In order to make the intensity vector have the same size as the logmel, it is calculated in the STFT domain and the mel space as

\begin{equation}
\begin{gathered}
    \boldsymbol{I}(f, t) = \frac{1}{\rho_{0} c} \Re\left\{\mathrm{W}^{*}(f, t) \boldsymbol{\cdot} \left[\begin{array}{c}{\mathrm{X}(f, t)} \\ {\mathrm{Y}(f, t)} \\ {\mathrm{Z}(f, t)}\end{array}\right]\right\}, \\
    \boldsymbol{I}^{\text{norm}}_{mel}(k, t) = -\boldsymbol{H}_{mel}(k,f)\frac{\boldsymbol{I}(f, t)}{\left\|\boldsymbol{I}(f, t)\right\|_{2}},
\end{gathered}
\end{equation}
where, $\rho_{0}$ and $c$ are the density and velocity of the sound, $\mathrm{W}, \mathrm{X}, \mathrm{Y}, \mathrm{Z}$ are the STFT of $\mathrm{w}, \mathrm{x}, \mathrm{y}, \mathrm{z}$, respectively, $\Re\left\{\cdot\right\}$ indicates the real part, $^*$ denotes the conjugate, $\|\cdot\|_{2}$ is a vector's \(\ell_{2}\) norm, $k$ is the index of the mel bins, and $\boldsymbol{H}_{mel}$ is the mel-band filter banks. In this paper, the three components of the intensity vector are taken as three additional input channels for the network.

\subsection{Network architecture}
\label{ssec:network}

\begin{figure*}[ht]
  \centering
  \scalebox{.9}{\centerline{\includegraphics[width=\textwidth]{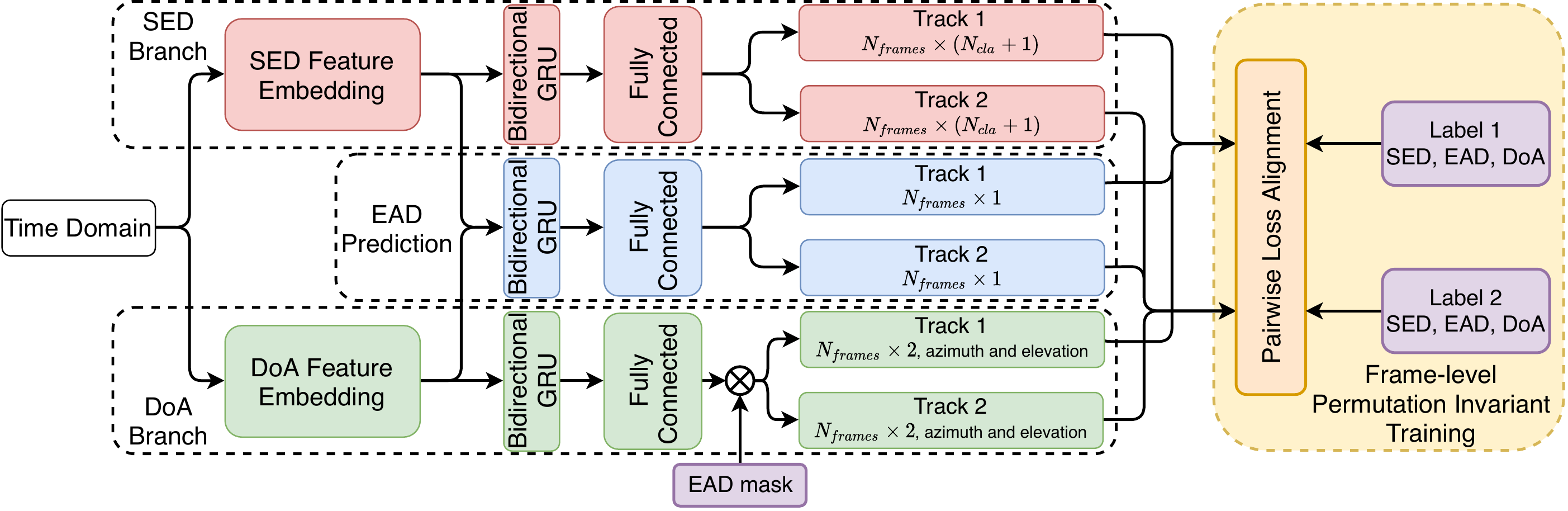}}}
  \caption{Network Architecture. $N_{frames}$ is the number of frames. $N_{cla}$ is the number of classes of events. In the SED branch, there is one additional class of event that is silence. EAD mask block is decribed in Eq. \ref{eqn:mask}.}
  \label{fig:network_architecture}
\end{figure*}

The proposed event-independent network uses track-wise outputs. For track-wise predictions, it is needed to pre-determine the number of tracks according to the maximum number of overlapping events. These tracks are event-independent, which means the prediction on each track can be of any type of event.

The network has two branches of feature embeddings, the SED branch and the DoA branch. Its architecture is shown in Fig. \ref{fig:network_architecture}. FOA time-domain signals are used as inputs and are first fed into two branches. In both of the SED and the DoA branches, a 1-D convolutional layer is first used to extract logmel spectrograms and intensity vectors. They are then normalized by batch-normalization layers. For the SED feature embedding, four groups of convolutional blocks are used to extract the SED embedding. Each convolutional block contains two 2-D convolutional layers with a kernel size of 3x3, a batch-normalization layer, and an average-pooling layer. For the DoA feature embedding, a revised ResNet 18 with two 3x3 2-D convolutional layers as the stem-layer is used. The size of the feature maps is 512 in the last layer for both SED and DoA embeddings. These two branches are then used to generate three predictions: SED predictions, EAD predictions, and DoA predictions. For SED and DoA predictions, the respective feature embeddings are fed into a two-layer bidirectional GRU and a fully-connected layer. For EAD predictions, SED and DoA embeddings are combined and are fed into a similar GRU and fully-connected layer. Outputs of the network have a track-wise format. Each track has at most one SED, one EAD, and one DoA prediction.

For SED, predictions have $N_{cla}$ + 1 types of events. Here, $N_{cla}$ is the number of event classes. The additional class of event indicates the silent class (no event is happening). The softmax activation function is used after each track for SED. The corresponding loss for SED predictions can be written as

\begin{equation}
\label{eqn:sed_loss}
\begin{gathered}
    \ell^{\text{SED}}_{t} (n_{\text{track}}) = -\log \left[\frac{\exp{(y^{\text{SED}}_{t} [n_{\text{track}}, \text {class}_{t}]})}{\sum_{j \in \mathbf{J}} \exp (y^{\text{SED}}_{t} [n_{\text{track}}, j])}\right], \\
    \mathcal{L}^{\text{SED}} = \sum_{n \in N_{\text{track}}, t} \ell^{\text{SED}}_{t} (n) ,
\end{gathered}
\end{equation}

\noindent where $\ell^{\text{SED}}_{t}$ is the track-wise loss. $\mathcal{L}^{\text{SED}}$ is the total SED loss for updating the model. $y_{\text{SED}}$ denotes the output logits of the SED fully-connected layer, $t$ indicates the frame, $n_{\text{track}}$ is the track index, $\text{class}_{t}$ is the ground truth target at frame $t$, $N_{\text{track}}$ is the number of tracks, and $\mathbf{J}$ is the class set. It is a multi-class single-label problem for each track and a multi-class multi-label problem for all of the tracks.

EAD predictions need to combine SED and DoA embeddings. Binary cross entropy is used as the loss. The existence of EAD predictions is important for three reasons. First, it uses SED and DoA feature embeddings together to predict on-set and off-set information. In this way, the frame-level information does not solely depend on a single branch, label information from both branches can be utilized. In the mean time, the EAD loss can be back propagated to affect both branches of the feature embeddings; second, it constrains SED and DoA feature embeddings to unify track-binding so that tracks do not permute in different branches. That is, track 1 in the SED prediction can always be tied to track 1 in the EAD and DoA predictions. Third, EAD predictions contribute to mask out DoA predictions.

For DoA, predictions for tracks contain azimuth and elevation angles. A linear activation function is used. Under these circumstances, when there is no event happening, the ground truth DoAs should not be zero. It is therefore reasonable to use a mask to shield those invalid frames. During training, ground truth EAD labels are used as the mask to filter out those frames with actual events happening. Whereas during test, an intersection set of the SED and EAD masks is used. The loss for DoA predictions can be written as Eq. \ref{eqn:doa_loss},

\begin{equation}
\label{eqn:doa_loss}
\begin{gathered}
    \ell^{\text{DoA}}_{t} (n_{\text{track}}) = \frac{1}{2} \sum_{\text{azi, elev}} \{\lVert y^{\text{DoA}}_{t} - \hat{y}^{\text{DoA}}_{t} \rVert_{p} \boldsymbol{\cdot} \pazocal{M}^{\text{EAD}}_{t} (n_{\text{track}})\}, \\ 
    \mathcal{L}^{\text{DoA}} = \frac{1}{\sum_{n \in N_{\text{track}}, t} \pazocal{M}^{\text{EAD}}_{t} (n)} \sum_{n \in N_{\text{track}}, t} \ell^{\text{DoA}}_{t}(n) , 
\end{gathered}
\end{equation}

\noindent where

\begin{equation}
\label{eqn:mask}
\scalebox{0.9}{$ \pazocal{M}^{\text{EAD}}_{t} = 
    \begin{dcases*}
        \hat{y}^{\text{EAD}}_{t} & for training, \\
        \begin{aligned}
            &\mathbbm{1}[y^{\text{EAD}}_{t} > \tau^{\text{EAD}}] \cap \\
            &\mathbbm{1}[y^{\text{SED}}_{t} = \max_{\mathbf{J}} y^{\text{SED}}_{t}][0:N_{cla}]
        \end{aligned}
        & for test,
    \end{dcases*} $}\\
\end{equation}

\noindent is the mask for DoA predictions. $\ell^{\text{DoA}}_{t}$ and $\mathcal{L}^{\text{DoA}}$ are defined the same as the SED loss. $\hat{y}^{\text{DoA}}_{t}$ is the DoA ground truth. $\lVert \cdot \rVert_{p}$ is the $p$-norm. $\hat{y}^{\text{EAD}}_{t}$ is the EAD ground truth. $\mathbbm{1}[\pazocal{C}]$ is the binarize function, where $\mathbbm{1}[\pazocal{C}]$ is 1 when $\pazocal{C}$ is True, and 0 when $\pazocal{C}$ is False. $\tau^{\text{EAD}}$ is the threshold for EAD. $\mathbf{J}$ is the class set. $[0:N_{cla}]$ means to take the first $N_{cla}$ items.

\subsection{Permutation Invariant Training}
\label{ssec:pit}

After track-wise predictions are obtained, frame-level permutation invariant training (tPIT) is adopted to tackle the track permutation problem. In Fig. \ref{fig:network_architecture}, the tPIT contains the Pairwise-Loss-Alignment block, which assigns labels to different tracks to constitute all of the possible combinations of prediction-label pairs. Then the total loss for each combination is calculated. The lowest total loss is selected as the actual loss to perform the back-propagation. Assume all of the possible combinations of prediction-label pairs constitute a permutation set $\mathbf{P}$. $\alpha(t) \in \mathbf{P}$ is one of the possible permutation pairs at frame $t$. The tPIT loss can be written as Eq. \ref{eqn:tpit},

\begin{equation}
\label{eqn:tpit}
    \pazocal{L}^{t P I T}_{t}=\min _{\alpha(t) \in \mathbf{P}} \sum_{ N_{\text{track}} } \{\ell^{\text{SED}}_{t, \alpha(t)} + \ell^{\text{EAD}}_{t, \alpha(t)} + \ell^{\text{DoA}}_{t, \alpha(t)} \},
\end{equation}

\noindent where $\ell^{\text{SED}}_{t, \alpha(t)}$ and $\ell^{\text{DoA}}_{t, \alpha(t)}$ are defined in Eq. \ref{eqn:sed_loss} and \ref{eqn:doa_loss}, respectively.  $\ell^{\text{EAD}}_{t, \alpha(t)}$ is the binary cross entropy loss. 

Therefore, the process of tPIT is not only to perform the classification or regression training but also to pair the most probable predictions and labels inside the network. In this way, the event-independent track permutation problem can be elegantly solved.

\section{Experiments}
\label{sec:experiments}

In this section, the experimental results using the proposed method on the TAU-NIGENS Spatial Sound Events 2020 dataset are described.

\subsection{Experimental Setup}
\label{ssec:exp_setup}

The dataset contains 700 sound event samples spread over 14 classes. Within the dataset, realistic spatialization and reverberation through RIRs were collected in 15 different enclosures. From about 1500 to 3500 possible RIR positions across different rooms were applied. The dataset contains both static reverberant and moving reverberant sound events. More information can be obtained in \cite{politis2020dataset, DCASE2020Task3}.

The evaluation metrics used consider the joint nature of localization and detection \cite{Mesaros_2019_WASPAA}. There are two metrics for SED, F-score $\left(F_{\leq T^{\circ}}\right)$ and Error Rate $\left(E R_{\leq T^{\circ}}\right)$. They consider true positives predicted under a distance threshold $T=20^{\circ}$ from the reference. For the localization part, there are other two metrics which are  classification-dependent. The first is a localization error $L E_{\mathrm{CD}}$ expressing average angular distance between predictions and references of the same class. The second is a localization recall metric $L R_{\mathrm{CD}}$ expressing the true positive rate of how many of these localization estimates were detected in a class, out of the total class instances.

To generate the weights of the 1-D convolutional layer for feature extraction. A \num{1024}-point Hanning window with a hop size of \num{480} points is used. Audio clips are segmented to have a fixed length of \num{5} seconds with \SI{80}{\%} overlap for training. The learning rate is set to \num{0.0005} for the first \num{60} epochs and is adjusted to \num{0.0001} for each epoch that follows. The final results are calculated after \num{80} epochs. A threshold of \num{0.5} is used to binarize EAD predictions.

\subsection{Comparison Systems and Results}
\label{ssec:comp_sys}

In order to assess the performance of the proposed system, an ablation study is performed. Several systems are compared, including:

\begin{itemize}[itemsep=0.05ex, leftmargin=0.3cm, label=\raisebox{0.25ex}{\tiny$\bullet$}]
    \item \textbf{Baseline-FOA}: The baseline method using Ambisonic.
    \item \textbf{Baseline-Mics}: The baseline method using microphone array.
    \item \textbf{Track-Wise 1}: Track-wise output without EAD and tPIT.
    \item \textbf{Track-Wise 2}: Track-wise output with EAD but without tPIT. SED predictions are used as the mask to filter out active tracks.
    \item \textbf{Track-Wise 3}: Track-wise output with EAD but without tPIT. SED and EAD predictions are together used as the mask to filter out active tracks according to Eq. \ref{eqn:mask}.
    \item \textbf{Event-Ind}: The proposed event-independent system with EAD and tPIT. SED and EAD predictions are together used as the mask.
\end{itemize}


All of the experimental results were evaluated on fold \num{1} and were averaged with 5 different trials. Results for comparison are shown in Fig. \ref{fig:performance}. It can be seen that all of the proposed methods are better than the baselines with two-stage method \cite{cao2019polyphonic}. A simple comparison for ablation study shows that without EAD and tPIT, the ``Track-Wise 1'' method is the worst among proposed methods. There seems to be an exception of $L E_{\mathrm{CD}}$, where ``Track-Wise 1'' method is even lower than ``Track-Wise 2'' and ``Track-Wise 3''. This is due to the reason that during experiments, it was found that there is a trade-off between $L R_{\mathrm{CD}}$ and $L E_{\mathrm{CD}}$. That is, when $L R_{\mathrm{CD}}$ is increased, $L E_{\mathrm{CD}}$ gets worse. This is probably because when more DoAs are detected, a higher number of wrong frame-wise DoA angles are predicted on average, hence $L E_{\mathrm{CD}}$ gets worse. ``Track-Wise 3'' is slightly better than ``Track-Wise 2'', which indicates that the mask using SED and EAD predictions is more effective than using SED preidictions alone. This additionally proves the significance of EAD.


\begin{figure}[ht]
  \centering
  \scalebox{1.}{\centerline{\includegraphics[width=\columnwidth]{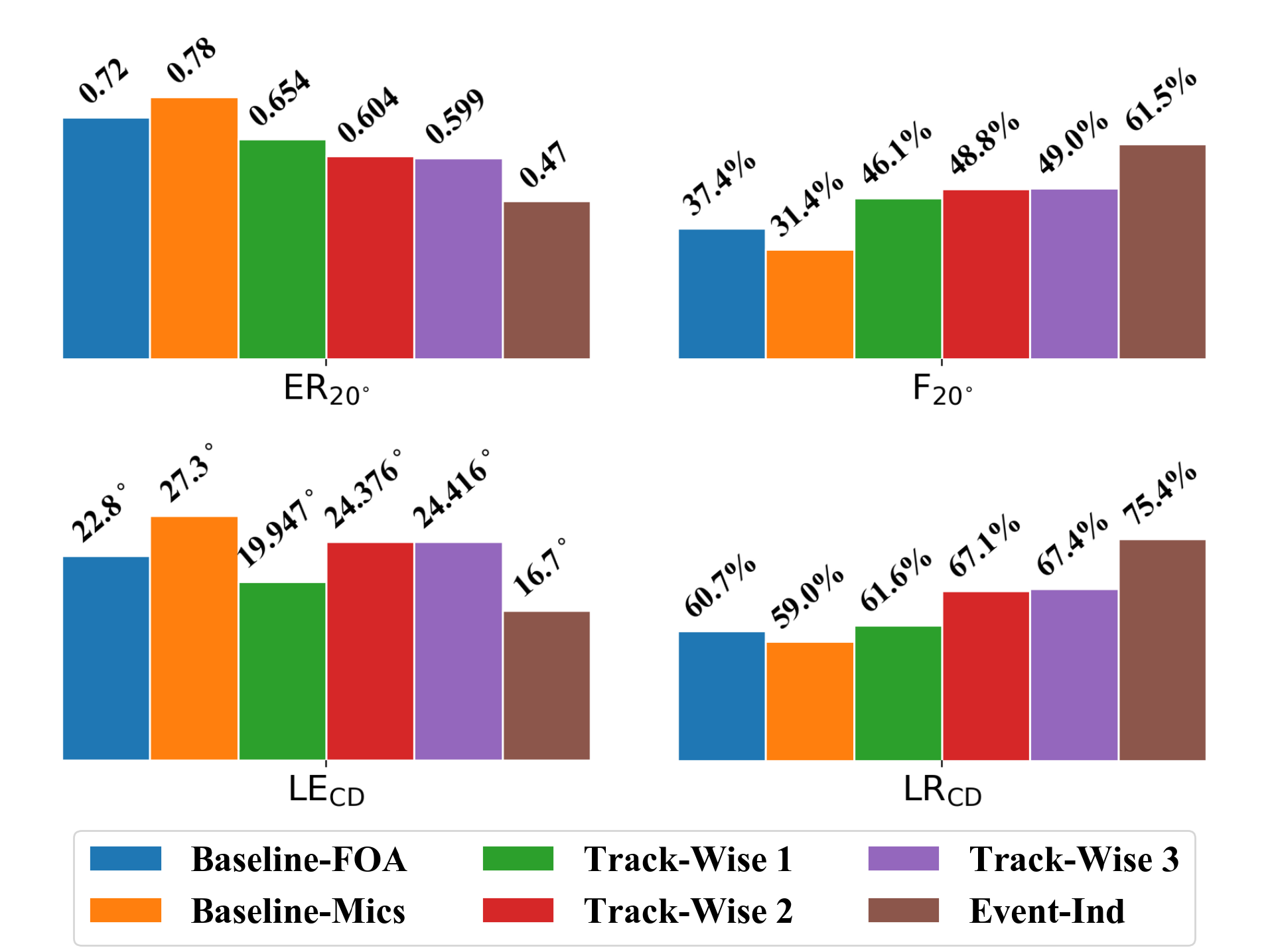}}}
  \caption{Comparison of different methods.}
  \label{fig:performance}
\end{figure}

The ``Event-Ind'' method achieves the best performance, which means additional EAD and tPIT features all contribute to increase the performance. The additional EAD prediction can constrain and unify predictions from SED and DoA branches both in terms of the temporal information and track-binding. The tPIT can rectify those incorrect label assignments and greatly increase the performance.


\section{Conclusion}
\label{sec:conclusion}
We proposed a new end-to-end event-independent network for polyphonic sound event localization and detection. The network treats polyphonic cases as multi-track problems, with each track having at most one event and its corresponding direction-of-arrival. In order to solve the problem of track permutation, a frame-level permutation invariant training strategy is employed. The network outputs three predictions which are sound event detection, event activity detection, and direction-of-arrival. Event activity detection encompasses the feature embedding information from both SED and DoA, hence is able to predict on-set and off-set times of events more accurately. The proposed system is easy to extend to more than two overlapping-event cases. Experimental results show that the proposed system outperforms the baseline methods by a large margin.

\section{ACKNOWLEDGMENT}
\label{sec:ack}
This work was supported in part by EPSRC Grants EP/P022529/1, EP/N014111/1 "Making Sense of Sounds", EP/T019751/1 "AI for Sound", National Natural Science Foundation of China (Grant No. 11804365), EPSRC grant EP/N509772/1, “DTP 2016-2017 University of Surrey”, and China Scholarship Council (No. 201906470002).

\bibliographystyle{IEEEtran}
\bibliography{refs}

\end{sloppy}
\end{document}